\documentclass[conference]{IEEEtran}
\IEEEoverridecommandlockouts
\usepackage{cite}
\usepackage{amsmath,amssymb,amsfonts}
\usepackage{algorithmic}
\usepackage{siunitx-v2}
\usepackage{graphicx}
\usepackage{textcomp}
\usepackage{xcolor}
\usepackage{algorithm,algorithmic}

\DeclareMathOperator {\atanz}{atan2}
\DeclareMathOperator {\asin}{asin}

\newcommand{\R}{\mathbb{R}}

\begin{document}

\title{Quality of Service Based Radar Resource Management for Navigation and Positioning}

\author{\IEEEauthorblockN{Tobias Müller, Sebastian Durst, Pascal Marquardt and Stefan Brüggenwirth}
\IEEEauthorblockA{\textit{Fraunhofer Institute for High Frequency Physics and Radar Techniques FHR}\\
Wachtberg, Germany \\
tobias.mueller@fhr.fraunhofer.de}
}

\maketitle

\begin{abstract}
In hostile environments, GNSS is a potentially unreliable solution for self-localization and navigation.
Many systems only use an IMU as a backup system, resulting in integration errors which can dramatically increase during mission execution.
We suggest using a fighter radar to illuminate satellites with known trajectories to enhance the self-localization information.
This technique is time-consuming and resource-demanding but necessary as other tasks depend on the self-localization accuracy.
Therefore an adaption of classical resource management frameworks is required.
We propose a quality of service based resource manager with capabilities to account for inter-task dependencies to optimize the self-localization update strategy.
Our results show that this leads to adaptive navigation update strategies,
mastering the trade-off between self-localization and the requirements of other tasks.
\end{abstract}

\begin{IEEEkeywords}
Q-RAM, Quality of Service, Radar, Navigation, Resource Management, Positioning, Cognitive Radar, GNSS
\end{IEEEkeywords}

\section{Introduction}
In a modern world, a global navigation satellite system (GNSS) is a simple and cost-effective solution for navigation purposes. However, it is insecure, as it can be spoofed and jammed, which is why in many cases alternative navigation solutions have to be available at least as a backup or for validation of the GNSS signals. Using only an  inertial measurement unit (IMU) is -- due to its drift and need for calibration -- insufficient in many long-term applications. For an airborne radar it seems natural to use its localization abilities to assist and calibrate the IMU on a regular basis.

Radar-based navigation is often done relative to the moving platform to solve local navigation problems \cite{kauffman2013real}.
To enable the use of radar as a global navigation sensor, targets with a known position are required. It is possible to use landmarks for that and then navigate with the help of synthetic aperture radar (SAR) images \cite{Lindstromnavi.533,veth2006fusion}. Another way is to use the reflections of moving targets with known trajectories.
In particular, the latter applies to satellites in low earth orbit.
They are highly available, while their density has recently been increasing heavily to enable internet services.
Additionally, the radar beam is pointing to the sky and thus the probability of intercept is reduced. Satellites are always available, whereas landmarks for SAR images may not be present offshore, in unknown terrain or after massive distortion of known terrain. Therefore satellites are used as a navigation aid throughout this paper.

Since modern multifunction radars have to accomplish many other tasks, it is a difficult problem to assign and schedule radar tasks for navigation updates in an optimal way.
This process lies in the responsibility of the resource manager, which has to select operational parameters from a multitude of possible task configurations differing in resource requirements and resulting utility with the goal to optimize the overall system performance under resource constraints.
This shows that the resource manager is an essential building block of any cognitive radar system (cf.~\cite{Haykin2006}).

A mathematical framework describing the problem is known as quality of service based resource allocation model (Q-RAM) \cite{Raj1997, Ghosh2006}.
Its classical solution strategy comprises pre-selecting beneficial task configurations by a convex hull operation in a resource-utility-space followed by an iterative global optimization scheme.
The standard Q-RAM framework is not able to model the effects of navigation update tasks appropriately and thus does not allow for a sophisticated decision-making. The reason is that classically, tasks are required to have an expected utility on their own.
We propose an evolution of this quality of service based method that instead calculates the utility of a navigation update using its impact on quality of the other tasks, thus contributing to the overall system utility (cf.~\cite{qran_sync_rww} for a related approach).

We use the radar simulation CoRaSi developed by Fraunhofer FHR to implement and showcase our new resource allocation scheme for navigation and compare it with a common rule-based scheduling method as well as with an alternative way of using Q-RAM in which regular navigation updates are scheduled.

The remainder of the paper is organized as follows.
Section~\ref{sec:resourcemanagement} gives a brief description of the two resource management principles used in later sections.
Section~\ref{sec:prob} introduces the resource management problem for navigation update tasks.
The framework used for the experimental verification is described in detail in Section~\ref{sec:framework}. Finally, Sections~\ref{sec:results} and~\ref{sec:conclusion} present the simulation results and the conclusion.

\section{Resource management}\label{sec:resourcemanagement}

This section introduces two different concepts of resource management (i.e.\ the prioritization, resource allocation and scheduling of tasks), which are used throughout the paper -- a traditional rule-based approach and Q-RAM.

\subsection{Rule-based resource management and scheduling}
A typical rule-based radar resource management is given by the time balanced scheduling \cite{butler_tracking_1998,miranda2004resource}. For this scheme a task comes with a due time and a priority number that defines a relation between different tasks. The scheduling now works as follows: In case at least one task is overdue, the overdue task with the highest priority is executed next. Otherwise, the task that is due next, will be executed. This way the scheduler automatically executes tasks near their due time.

In this framework, the allocation of resources (other than the decision if a task is executed at all) is performed using fixed rules. Exemplary rules are to use the desired probability of detection of a potential target at the instrumented range and the desired revisit time to specify a search task, or to use the desired track sharpness, i.e.\ the ratio of beam width and tracking uncertainty, to configure a track update task.

\subsection{Q-RAM model}
This section briefly introduces the Q-RAM model and its classical solution approach (cf.~\cite{Ghosh2006}).

Q-RAM assigns limited radar \emph{resources} like power, bandwidth, or computation time in an optimal manner to the tasks the system has to perform. This is done by selecting \emph{operational parameters} like integration times, waveforms or tracking filters, i.e.\ task configurations.
These configurations are evaluated with respect to certain performance measures, i.e.\ \emph{qualities}, which are also influenced by 
\emph{environmental conditions}, which include non-manipulable factors like maps, weather or targets. 
A simple example of a task quality is the inverse track error for a track update task. The quality can be used to implement hard requirements for specific functions of a system and additionally should be defined such that they are interpretable by a human operator.
To compare different qualities, encode mission goals and task priorities, a scalar value called \emph{utility} is associated with a task configuration's qualities and the existing environmental conditions.
Q-RAM then tries to maximize the sum of all utilities.

Mathematically, this can be formulated as follows (taken from \cite{qram_rl}).
Let $\{\tau_1,\ldots,\tau_n\}$ be a set of radar tasks and let there be $k$ types of resources with resource bounds $R_1,\ldots,R_k$.
Associated with each task $\tau_i$ are
\begin{itemize}
	\item a discrete operational space $\Phi_i$, i.e.\ a discrete space of feasible task configurations,
	\item a function $g_i\colon\thinspace \Phi_i\rightarrow\mathbb{R}^k$ mapping task configurations to their resource requirements,
	\item a quality space $Q_i$ and
	an environment space $E_i$,
	\item a map  $f_i\colon\thinspace \Phi_i\times E_i\rightarrow Q_i$ associating a quality level to a configuration-environment-pair and
	\item a quality-based utility function $\widetilde{u}_i\colon\thinspace Q_i\times E_i\rightarrow\mathbb{R}$.	
\end{itemize}
We define $u_i\colon\thinspace \Phi_i\times E_i\rightarrow\mathbb{R}$ via
$u_i(\phi, e) := \widetilde{u}_i(f_i(\phi, e), e)$
and the \emph{system utility} $u$ for chosen configurations
$\phi = (\phi_1,\ldots,\phi_n) \in \Phi := \Phi_1\times\cdots\Phi_n$
under environmental conditions
$e=(e_1,\ldots,e_n) \in E := E_1\times\cdots E_n$ as
$
u(\phi, e) = \sum_{i=1}^{n} u_i(\phi_i, e_i).
$
Now for fixed environmental data $e\in E$, the aim is to optimize global system utility while respecting resource bounds, i.e.\ we have 
the following optimization problem:
\begin{align}
	\begin{split}
	\max_{\phi = (\phi_1,\ldots,\phi_n)}& u(\phi, e)\\
	\textrm{s.t. } \forall j=1,\ldots,k\;\;& \sum_{i=1}^n \big(g_i(\phi_i)\big)_j \leq R_j.
	\end{split}
\end{align}

A solution to the Q-RAM problem is proposed in \cite{Raj1997, Lee1998, Lee1999, Ghosh2004, Ghosh2006} and will be briefly outlined in the following.
If there are multiple types of resources $R_1,\ldots,R_k$, a so-called \emph{compound resource} is used, i.e.\ a function
$h\colon\thinspace\mathbb{R}^k\rightarrow\mathbb{R}$ mapping a resource vector to a scalar measure of resource requirements.
On a per task basis, all possible task configurations are generated and evaluated. This yields an embedding from the space of task configurations into resource-utility-space. A convex hull operation is used to determine the subset of configurations maximizing utility for fixed resource levels.
A global optimizer then iteratively allocates resources to the task offering the best utility-to-resource-ratio provided sufficient resources are available.
After the resource allocation step, the resulting tasks are placed on the timeline by a scheduler.

\section{Problem definition and solution approach}\label{sec:prob}
This section further describes and investigates the resource management problem for navigation tasks and presents our proposed modification of the resource management framework to solve it.

Due to the high range of the satellites, using their radar reflections in navigation leads to a high additional resource request by the navigation task.
This resource demand is added to the requirements of the many other tasks a modern multifunctional radar system has to complete.
For our proof-of-concept, we use air-to-air track updates and surveillance as typical such radar tasks.
Traditionally, navigation updates would be executed according to a regular schedule (independent of the given situation) and with fixed heuristics for resource allocation, which increase the implementation complexity of the system.

Arguably, the challenges current and future radar systems have to face require cognitive approaches that include possibly irregular navigation update strategies that are adaptive to the environmental conditions, such as target behaviour and current system load.
As one feasible solution, we propose to enhance the resource manager with these capabilities. The performance of classical radar tasks is highly related to the accuracy of self-localization, thus leading to tight dependencies between the tasks.

As already stated, the classical Q-RAM idea does not refer to any dependency or influence between different tasks, beside their resource usage. The idea investigated here is now to include a functionality, like navigation, that may or may not directly have a utility. On an airborne platform, the detection of hostile targets in an area is less useful if an unknown error adds to the track from a viewpoint outside the carrier platform.

To use Q-RAM for our new approach, we perform the optimization for every navigation task configuration. The configurations for a navigation update differ in the satellites to be illuminated, in general including different integration times. This leads to different probabilities of detection and different expected errors.

Let the different possible configurations for a navigation update be denoted by $c_1,\ldots,c_n$. Each of these possesses a respective resource requirement $r_1,\ldots,r_n$ for the update.

The aim is to pick the configuration which yields the highest overall system utility. In the simplest form, this is the binary decision of performing ($c_1$) or not performing ($c_2$) a navigation task in a given planning period, in which case $r_2 = 0$.
For a given $c_i$ ($i\in\{1,\ldots,n\}$), we compute the self-localization quality and its impact on the expected utility of the other tasks which requires a non-trivial redesign of the utility functions.
Then we subtract the resource demand $r_i$ of the navigation task from the total resource budget and perform a Q-RAM optimization using these adapted utility functions.
Afterwards, the overall best case is chosen and executed.
This process is repeated in every planning interval, leading to a dynamic update pattern which takes into account the varying environmental conditions.

It is worth noting, that the computational load increases with the number of possible configurations of the navigation update but this can for example be mitigated by parallel execution as well as re-using the concave majorants of the tasks unaffected by the self-localization error.

\section{Experimental validation framework}\label{sec:framework}

This section describes in detail the scenario, simulation environment and models used in the experimental verification of the proposed method.

\subsection{Scenario}
\begin{figure}[htbp]
	\centering
	\includegraphics[width=0.45\textwidth]{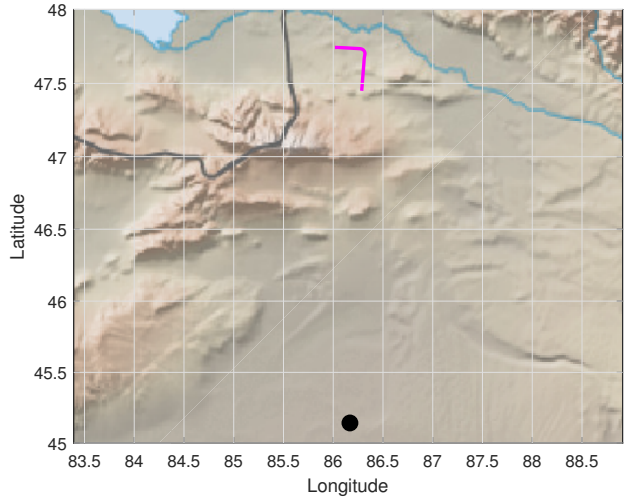}
	\caption{Trajectory of the plane (pink) and region of the targets (black).}\label{fig:scenario}
\end{figure}

The scenario is composed of three types of vehicles: a friendly fighter, several thousand Starlink satellites and four enemy air targets.
The friendly fighter carries a nose radar and its trajectory begins straight, where after \SI{300}{\s} a \SI{90}{\degree} right turn is initiated (see Figure~\ref{fig:scenario}).
This situation might occur when a queued surveillance volume is externally triggered, where undetected targets are expected.
After \SI{500}{\s} the scenario ends.
The Starlink satellites where chosen since their orbits are low and their density is high enough to expect many situations where multiple satellites are visible at the same time. The satellite data was extracted from \cite{kelso2023celestrak}.
Four air targets, with a radar cross section (RCS) of \SI{0.1}{\m\squared} each, have to be detected and tracked by the radar. Their position is chosen such that they become visible after the fighter performs its turn.

The fighter's initial position uncertainty (given in meters) is set to a random error of $\mathcal{N}(0, 100)$ for each axis and the velocity error begins analogously with $\mathcal{N}(0, 10)$.
The random initialization is chosen for two practical reasons. A perfectly known position at the beginning on the one hand means IMU integration errors stay relatively low for some time, thus unnecessarily increasing the length of the scenario. On the other hand, it is simply unrealistic in mid-flight.

\subsection{CoRaSi}\label{sec:corasi}
Fraunhofer FHR’s Cognitive Radar Simulator (CoRaSi) has been developed to investigate new radar resource management and tracking techniques for generic phased array radars. It enables the prediction of the performance of specific radar systems in various situations. This is specifically of interest when a radar has to be placed in a new situation or  
during the development of a new system when the performance requirements have to be determined.

The real-time analysis of radar-systems can be coupled with other real-time simulations such as flight-simulations. A modular implementation enhances fast development of new search patterns, tracking filters, association techniques, resource management algorithms and much more. In case that multiple sensors are simulated simultaneously, centralized, decentralized and hybrid fusion techniques are available. This allows a search radar for instance to cue an acquisition radar. The tracker is composed of the two levels: association and filtering. Implemented algorithms for association are based on global nearest neighbour and Gaussian mixture probability hypothesis density (GM-PHD) tracking \cite{yazdian2015refined}. Available tracking filters are extended, iterated extended, unscented Kalman filter (EKF, IEKF, UKF) and a particle filter \cite{blackman1999design,ZHANG20134468,ristic2003beyond}.

Target trajectories are available based on mathematical models which can be generated by CoRaSi on its own like constant velocity, singer or (manoeuvrable) ballistic models. An external continuously growing database is available as well, based on Automatic Dependent Surveillance-Broadcast (ADS-B), measurement campaigns or two-line elements (TLE) for satellites \cite{kelso2023celestrak} datasets.  Aspect angle dependent RCS values are used to calculate accurate signal-to-noise ratio (SNR) estimations, if available. The simulation can investigate propagation effects like refraction, atmospheric attenuation, ground clutter and ground reflection.

\begin{figure*}[ht]
	\centering
	\includegraphics[width=0.9\textwidth]{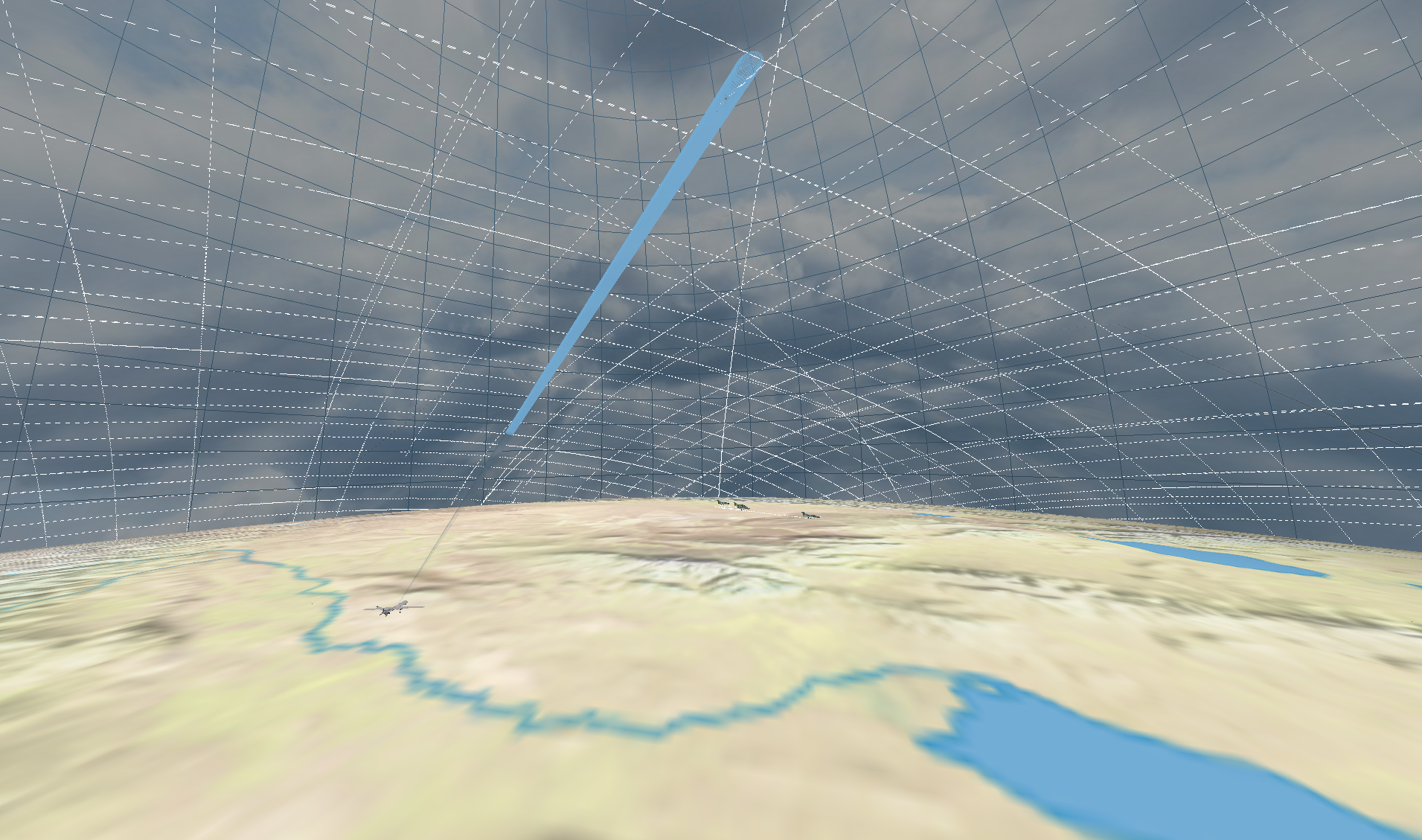}
	\caption{3D view of CoRaSi during a navigation beam update.}\label{fig:3dview}
\end{figure*}

\begin{figure}
	\centering
	\includegraphics[width=0.23\textwidth,height=0.23\textwidth]{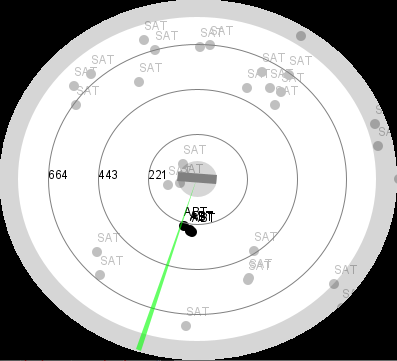}
	\includegraphics[width=0.23\textwidth,height=0.23\textwidth]{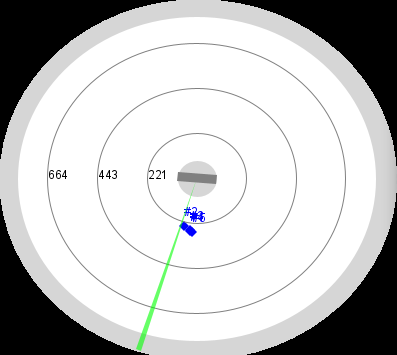}
	\caption{$2$-dimensional view of the ground truth (left) and the resulting tracking  (right), where satellites are intentionally not plotted.}\label{fig:2dview}
\end{figure}

\begin{figure}
	\centering
	\includegraphics[width=0.5\textwidth]{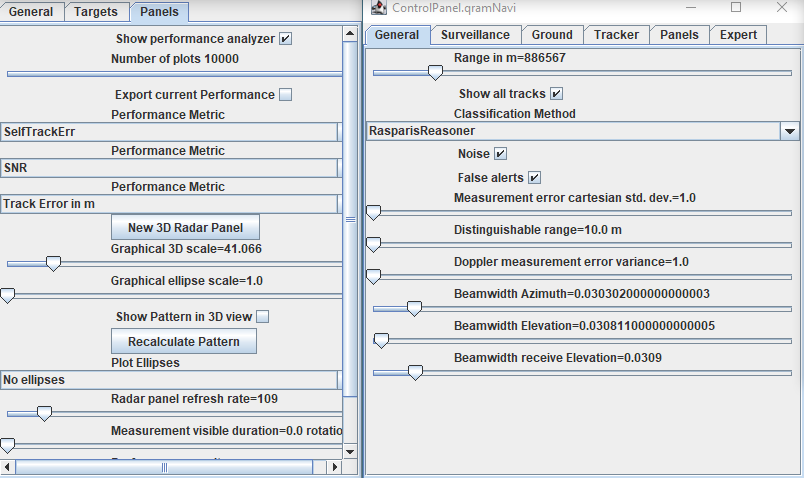}
	\caption{Exemplary settings for the ground truth and main panels (left) and radar settings (right). \label{fig:settings}}
\end{figure}

\begin{figure}
	\centering
	\includegraphics[width=0.5\textwidth]{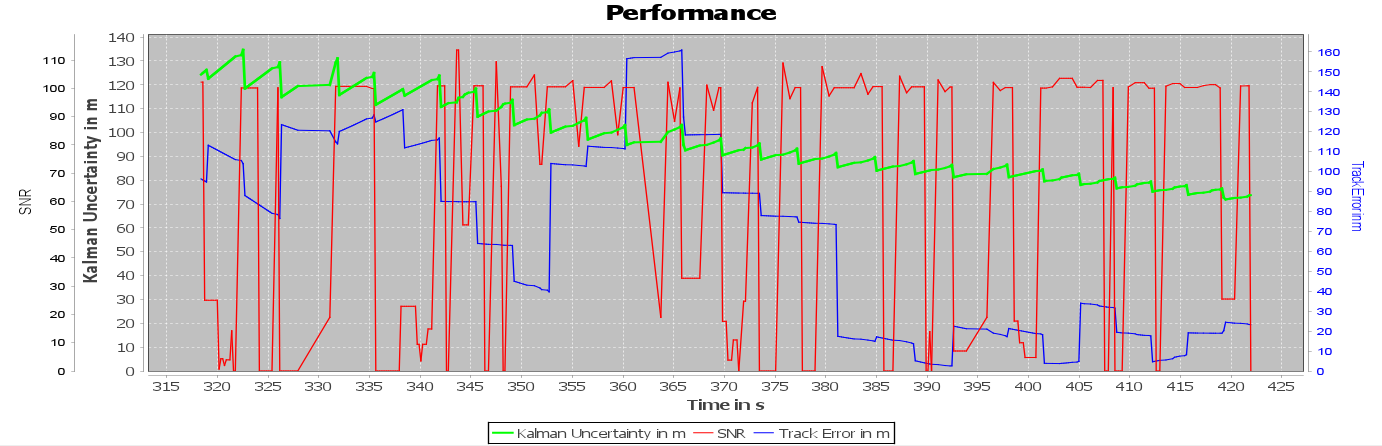}
	\caption{Visualization panel with several metrics as chosen in settings panel. }\label{fig:performance}
\end{figure}

CoRaSi includes an optional 3D view (Figure~\ref{fig:3dview}) and a GUI (Figures~\ref{fig:2dview} to~\ref{fig:performance}), which allows a visualization of the environment. CoRaSi was originally focused on ground-based platforms but moving vehicles are now investigated as well. CoRaSi is mainly written in Java and mostly developed as part of the basic funding by the German Ministry of Defence.


\subsection{Modelling}\label{sec:model}
This section describes the mathematical framework used to model an airborne radar using known object trajectories for navigation. A simplistic IMU model is used to infer measurements with high update rate. The association problem is neglected in this investigation.
\subsubsection{Navigation}
The IMU is simplified for demonstration purposes and to show the opportunities given by a radar-based positioning system. The orientation error is assumed to be negligible. The aircraft's position $p$ and velocity $v$ are calculated by integrating the acceleration $\widetilde{a}$ as measured by the IMU. Its measurement is assumed to be the true acceleration $a$ plus white Gaussian noise $\nu$. This is used as a process description for the extended Kalman filter:
\begin{align}
	\begin{split}
	x(t) &= \begin{pmatrix}  p(t) \\ v(t) \end{pmatrix}	\\
	\dot{x}(t) &= \begin{pmatrix}	0 & I \\ 
			0 & 0 \end{pmatrix} \, x(t) + \begin{pmatrix}	0 \\%
			I	\end{pmatrix} \, \widetilde{a}(t) , \quad \textrm{with identity }I \\
	\widetilde{a} &= a + \nu \quad \nu \in \mathcal{N}(0,\sigma_\text{IMU})
	\end{split}
\end{align}

To be able to implement the extended Kalman filter, the coordinate transformation from measurement space to process space for each satellite is necessary.
The measurement space consists of azimuth, elevation and range which together form spherical coordinates relative to the antenna position and orientation. For every instance of time the Earth-centred, Earth-fixed (ECEF) position of all satellites is assumed to be perfectly known. The transformation of some estimated position of the vehicle $p \in \R^3$ to a point $(\theta, \phi, R)^T \in \R^3$ in measurement space using the known position of the satellite is defined via
\begin{align}
	\begin{split}
	c_\text{NED} &= T_\text{NED}^p (p_i), \\
	c = (c_x,c_y,c_z)^T &= R(\alpha, \beta, \gamma) c_\text{NED},	\\
	\begin{pmatrix}\theta \\ \phi \\ R \end{pmatrix}
	&= \begin{pmatrix}
		\atanz({c}_x,{c}_x)	\\
		\asin(-{c}_z/||c ||_2)	\\
		|| c ||_2
	\end{pmatrix},	
	\end{split}
\end{align}
where $(\alpha, \beta, \gamma) \in \R^3$ define the Eulerian angles of the antenna / vehicle based on the north, east, down (NED) coordinate system for an antenna that has no mechanical steering capability. $R(\alpha, \beta, \gamma)$ is the rotation matrix defined by the Eulerian angles and $T_\text{NED}^p$ transforms ECEF to NED coordinates \cite{zhang2012sensor}.

The Jacobian matrix of the coordinate transformation for the EKF is calculated using a numerical scheme.

\subsubsection{Radar model}
Radars are sensors that produce measurements of a target when enough of the transmitted energy is received and integrated. Typically, the measurements are in spherical coordinate space (azimuth, elevation, range) and range rate. Modern nose radars of fighters have many antenna elements with individual phase control to be able to steer the beam electronically.

Several mathematical models approximate the behaviour of radars. The one used for the simulation discussed in this article are described in the following.
The SNR after matched filter integration is given by the radar equation \cite{mahafza2013radar}
\begin{align}\label{eq:radareq}
	\text{SNR} = \frac{P G^2 \tau_d \lambda^2 L }{(4 \pi)^3 R^4 k T n_f L_\text{air}},
\end{align}
where $P$ is the peak power emitted by the antenna, $G$ is the gain, which is assumed to be symmetric in transmit and receive, $\tau_d$ gives the total time the radar is transmitting, and $\lambda$ is the wavelength at the centre frequency, which is typically around \SI{10}{\GHz}.
The additional system losses are denoted by $L$ and $n_f$ is the noise factor of the receiver amplifier chain.
$L_\text{air}$ includes the attenuation due to the waves travelling through air, $k$ is the Boltzmann constant and $T$ gives the receivers noise temperature.
The gain from equation \eqref{eq:radareq} for an array with $N_\theta \times N_\phi$ elements is defined as
\begin{align}\label{eq:gain}
	&G(\theta,\bar{\theta},\phi,\bar{\phi}) = G_{\text{bs}} \, \text{AF} (\theta, \bar{\theta}) \text{AF} (\phi, \bar{\phi}) 
\end{align}	
with the array factor
\begin{align}
	\text{AF}(\star, \bar{\star}) &= \frac{ |\sum_{k=0}^{N_\star} \exp{\left\{ i \frac{2 \pi k}{\lambda} d_\star (cos(\star)-cos(\bar{\star})) \right\} }| }
	{N_\star^2} 
\end{align}
for $\star \in \{ \theta, \phi \}$,
which calculates the steering losses and the beam broadening effect due to electronic steering. The steering angles are given as $\bar{\theta}, \bar{\phi}$ and the evaluation angles as $\theta, \phi$, respectively.
The broadside gain is denoted by $G_{\text{bs}}$, which is the highest possible value of~\eqref{eq:gain}.
Note that the calculation assumes a rectangular array, where $d_\star$ denotes the distance between the antenna elements in the direction of azimuth $\theta$ or elevation $\phi$.

The measurement accuracy is modelled as in \cite{blackman1999design}, neglecting the bias and hardware limitations in high SNR cases, since the satellites will not be illuminated with that much energy:
\begin{align}
	\sigma_\star &= \frac{0.628 \, \star}{2 \sqrt{\text{SNR}}},  \\
	\sigma_r &= \frac{c}{\sqrt{12} \, B},
\end{align}
with $c$ the speed of light and $B$ the bandwidth.

\begin{figure}[htbp]
	\centering
	\includegraphics[width=0.35\textwidth]{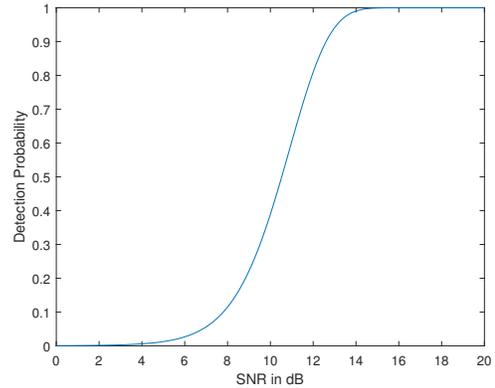}
	\caption{Detection probability in dependence of the signal-to-noise ratio for false alarm probability $10^{-6}$
		(see \cite{van2004detection}).}\label{fig:pd}
\end{figure}

The duration of an executed task is proportional to the expected SNR, which correlates with the probability of detecting a target or satellite (see Figure~\ref{fig:pd}).
Detections are simulated according to this probability and only if a task results in a detection, it is considered as successful.

The assumed radar data are documented in Table \ref{tab:radar}.

\begin{table}[htbp]
	\caption{Radar parameters}\label{tab:radar}
	\begin{center}
		\begin{tabular}{lr}
			\hline
			\textbf{Name} & \textbf{\textit{Value}} \\
			\hline
			frequency & \SI{10}{\GHz}	\\
			peak power & \SI{10}{\kW}	\\
			antenna elements & $60 \times 60$	\\
			maximum pulse duration & \SI{590}{\micro \s}	\\
			duty cycle & \SI{10}{\percent}	\\
			bandwidth & \SI{86}{\MHz}	\\
			gain & \SI{42}{\dB}	\\
			receiver noise temperature & \SI{330}{\K}	\\
			noise factor & \num{2}	\\
			additional losses & \num{0.5}	\\
			\hline
		\end{tabular}
		\label{tab1}
	\end{center}
\end{table}

\subsubsection{Resource management implementation}
We investigate three resource management solutions. Firstly, the traditional time balanced scheduling, secondly, Q-RAM with fixed navigation update intervals, and thirdly, the modified Q-RAM considering the influence of different navigation update times.

For the traditional time balanced scheduling and Q-RAM with regular navigation updates a rule is necessary that defines the resource management part. The used rule consists of two steps. The first step is to choose a satellite from the database. From all visible satellites, one maximizing the expression  
\begin{align}
	\frac{p^T P p}{||p||_2^2}
\end{align}
is chosen,
where $P$ is the estimated error covariance matrix of the navigation Kalman filter and $p$ the respective satellite's position.
Since the range measurement of a radar is much better than its angular measurements, this rule leads to switching between satellites and thus reducing all error dimensions after multiple measurements.
In the second step a navigation update task is generated with coherent integration time high enough to reach an SNR of \SI{13}{\dB}. Since at least three different satellite measurements are necessary to reach a high quality position measurement, the presented rule is repeated up to three times (provided there are at least so many satellites visible within this short time interval).

For the improved Q-RAM scheme, the impact of the navigation tasks on other tasks has to be estimated. Tracking quality is defined by the inverse of the determinant of the estimated error covariance at the end of the planning interval (here chosen to be \SI{10}{\s}), which is a measure for the estimated track error.
The mathematical models of the quality and utility functions are as follows.
For a tracking task, we have
\begin{align}
	q_\textrm{track} &= \frac{10}{ \det(P)^{\frac{1}{6}}} \in \R
\end{align}
and
\begin{align}
	u_\textrm{track} &= (1 - e^{-10q_\textrm{track}})/10,
\end{align}
where $P$ is the error covariance matrix for the track including the navigation error, but propagated to the end of the planning interval including all potential measurements.
For the search task, the functions are given by
\begin{align}
	q_\textrm{search} &= (R, \frac{10}{\det(P')^{\frac{1}{6}}})^T \in \R^2
\end{align}
and
\begin{align}
	\begin{split}
	u_\textrm{search} &= 0.01 \, (q_\textrm{search})_1+0.99 \, \min((q_\textrm{search})_2,1) \\
	& = 0.01 \, R +0.99 \, \min\left(\frac{10}{\det(P')^{\frac{1}{6}}}, 1\right),
	\end{split}
\end{align}
where $P'$ denotes the expected position error covariance matrix of the navigation filter after executing or not executing, respectively, the navigation task and $R$ is the resource allocated to the surveillance task. This means that its quality increases, if it is given more time on the antenna.

\subsubsection{Trajectories}

The airborne trajectory of the fighter is generated using FlightGear \cite{fg2023} and its autopilot. This is set to a height of \SI{1830}{\m} and a velocity of \SI{113}{\m \per \s}. Initially, the heading is set to \SI{90}{\degree} and after \SI{190}{\s} it is set to \SI{180}{\degree} to fulfil the turn. The trajectory is logged to a file such that the exact same trajectory is run for the different resource managers under consideration. The trajectory is visualized in Figure~\ref{fig:scenario} containing the region of four targets with continuous velocities of up to \SI{300}{\m \per \s}.

The satellite TLEs are downloaded from \cite{kelso2023celestrak}. This data is used to generate files that contain a \SI{500}{\s} long trajectory that is already converted to ECEF, which improves simulation speed.

\section{Results}\label{sec:results}
This section presents the comparison of the proposed method denoted by \emph{Q-RAM~nav} with the regular Q-RAM and time balanced schemes for fixed navigation update intervals of \SIlist{10; 20; 30}{\s}, denoted by \emph{Q-RAM~10}, \emph{Q-RAM~20}, \emph{Q-RAM~30}, \emph{TB~10}, \emph{TB~20}, \emph{TB~30}, respectively.
Every algorithm has been tested in 18 runs generated by a Monte Carlo simulation as a variation of the base scenario described in Section~\ref{sec:framework}.

\begin{figure}[htbp]
	\centering	
	\includegraphics[width=0.23\textwidth,height=0.23\textwidth]{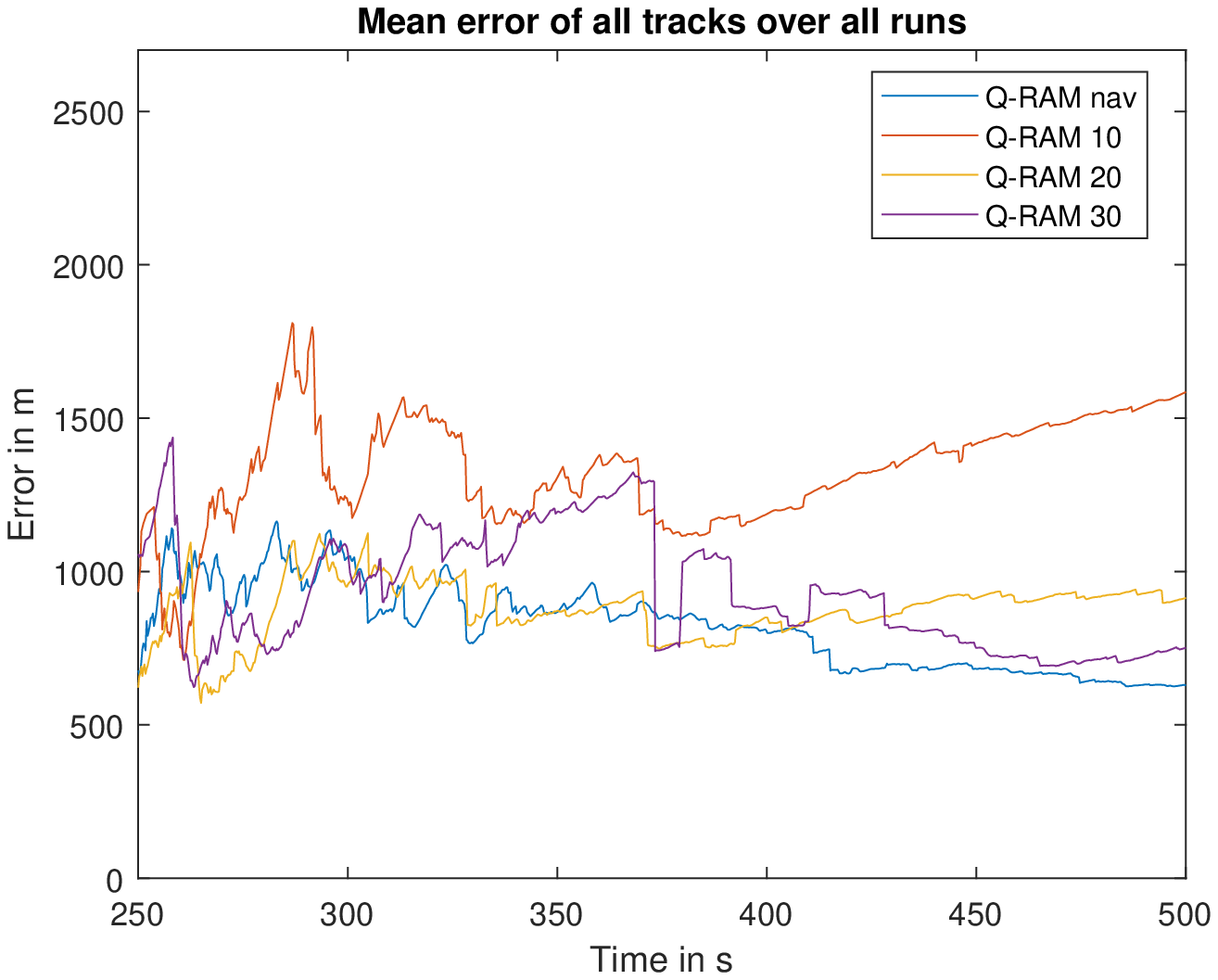}
	\includegraphics[width=0.23\textwidth,height=0.23\textwidth]{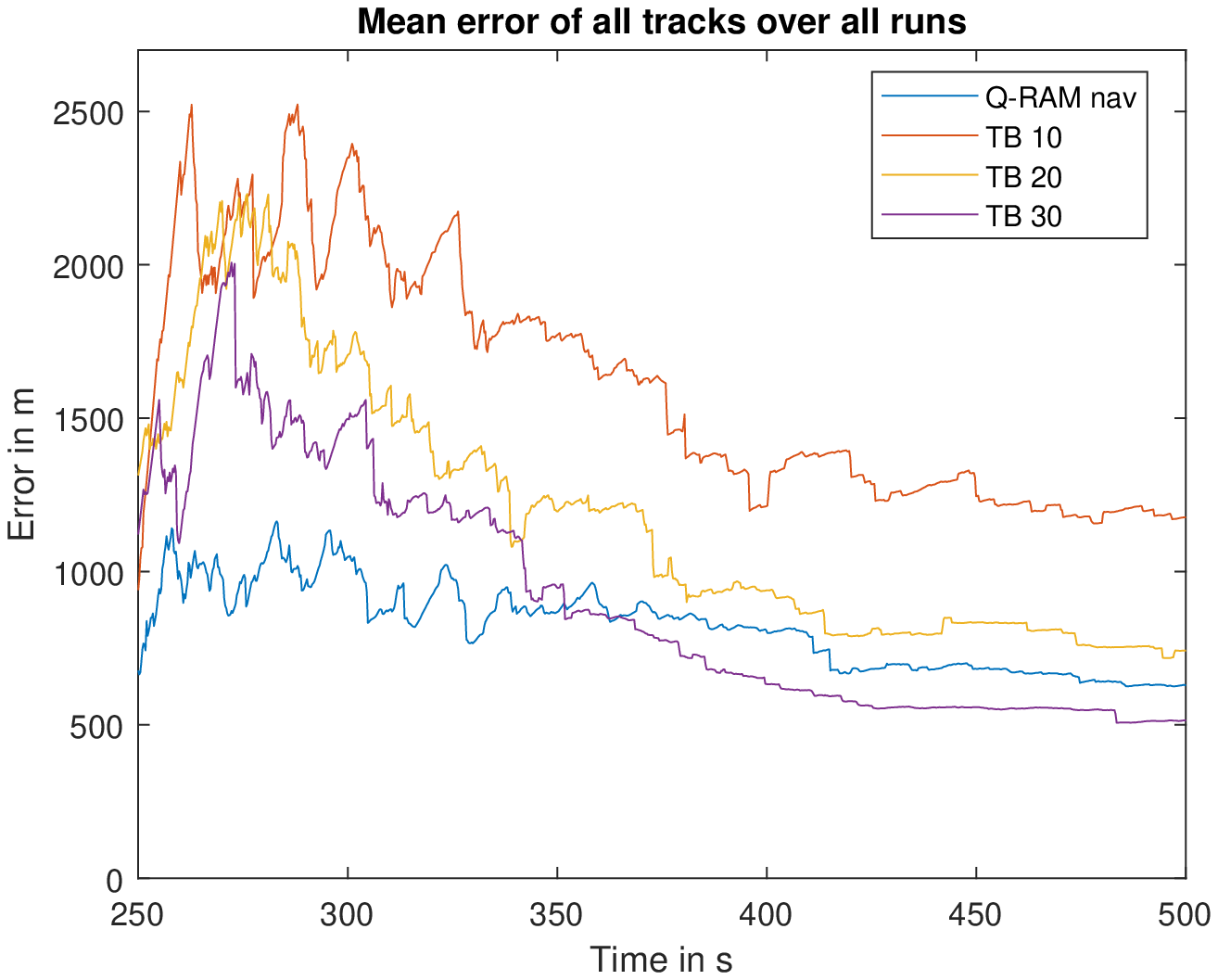}
	\caption{The mean errors for all tracks and runs are shown for the different Q-RAM schemes (left) and the time balanced techniques (right).}\label{fig:allres}
\end{figure}

\begin{table}[htbp]
	\caption{Mean track errors and standard deviation in meters}\label{tab:trackerror}
	\begin{center}
		\begin{tabular}{lrr}
			\hline
			 & \textbf{mean} & \textbf{std.\ dev.} \\
			\hline
			 \emph{Q-RAM nav} & 822  & 136 \\
			 \emph{Q-RAM 10} & 1328 & 176 \\
			 \emph{Q-RAM 20} & 878 & 94 \\
			 \emph{Q-RAM 30} & 929 & 188  \\
			 \emph{TB 10} & 1611 & 386 \\
			 \emph{TB 20} & 1183 & 422 \\
			 \emph{TB 30} & 914 & 389 \\
			\hline
		\end{tabular}
		\label{tab1}
	\end{center}
\end{table}

The mean track errors over all \num{18} runs for all tracks are shown in Figure~\ref{fig:allres} and Table~\ref{tab:trackerror}.
The overall tracking error of \emph{Q-RAM nav} is best and its stability on par with the best baseline algorithms.

\begin{figure}[htbp]
	\centering	
	\includegraphics[width=0.23\textwidth,height=0.23\textwidth]{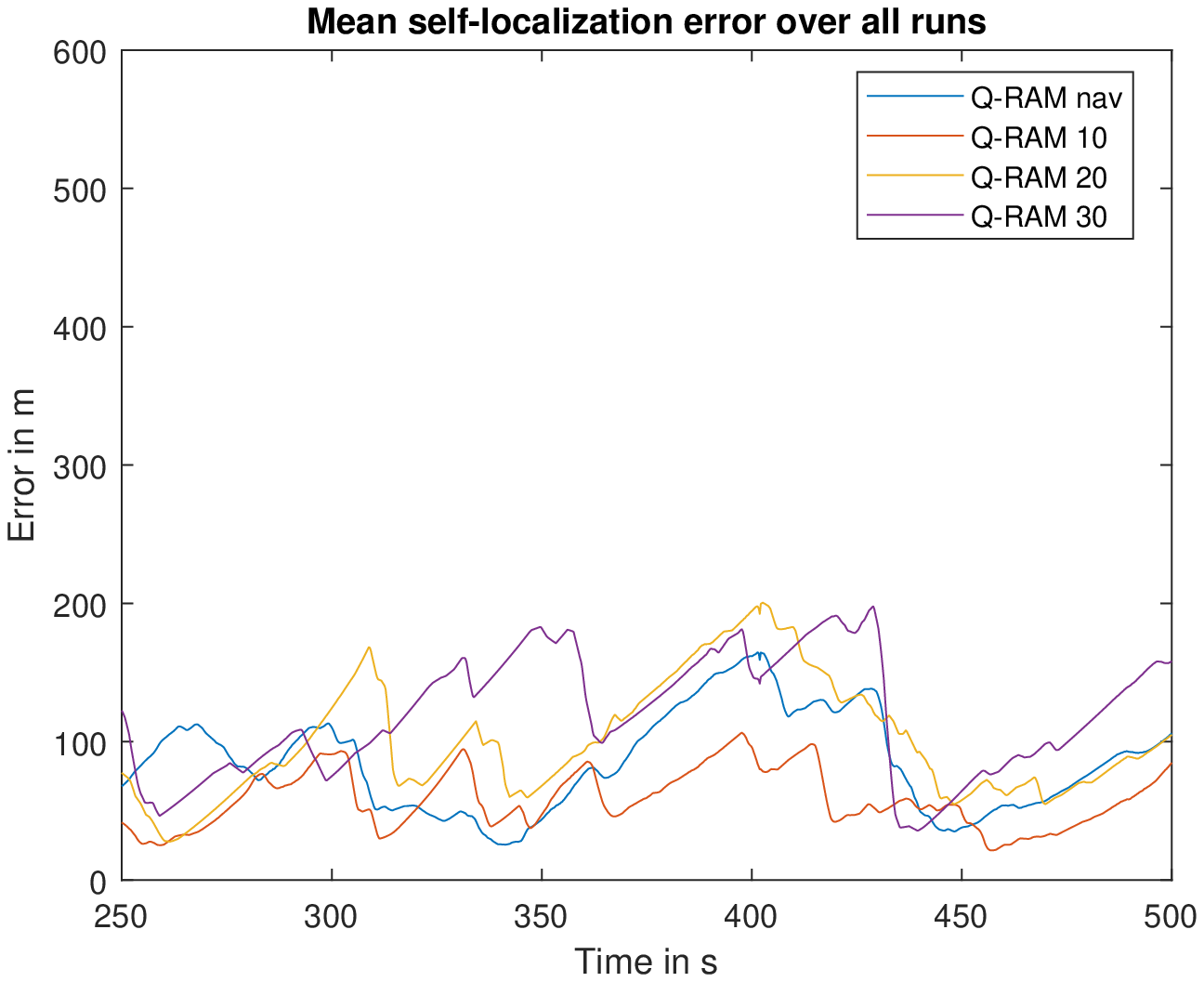}
	\includegraphics[width=0.23\textwidth,height=0.23\textwidth]{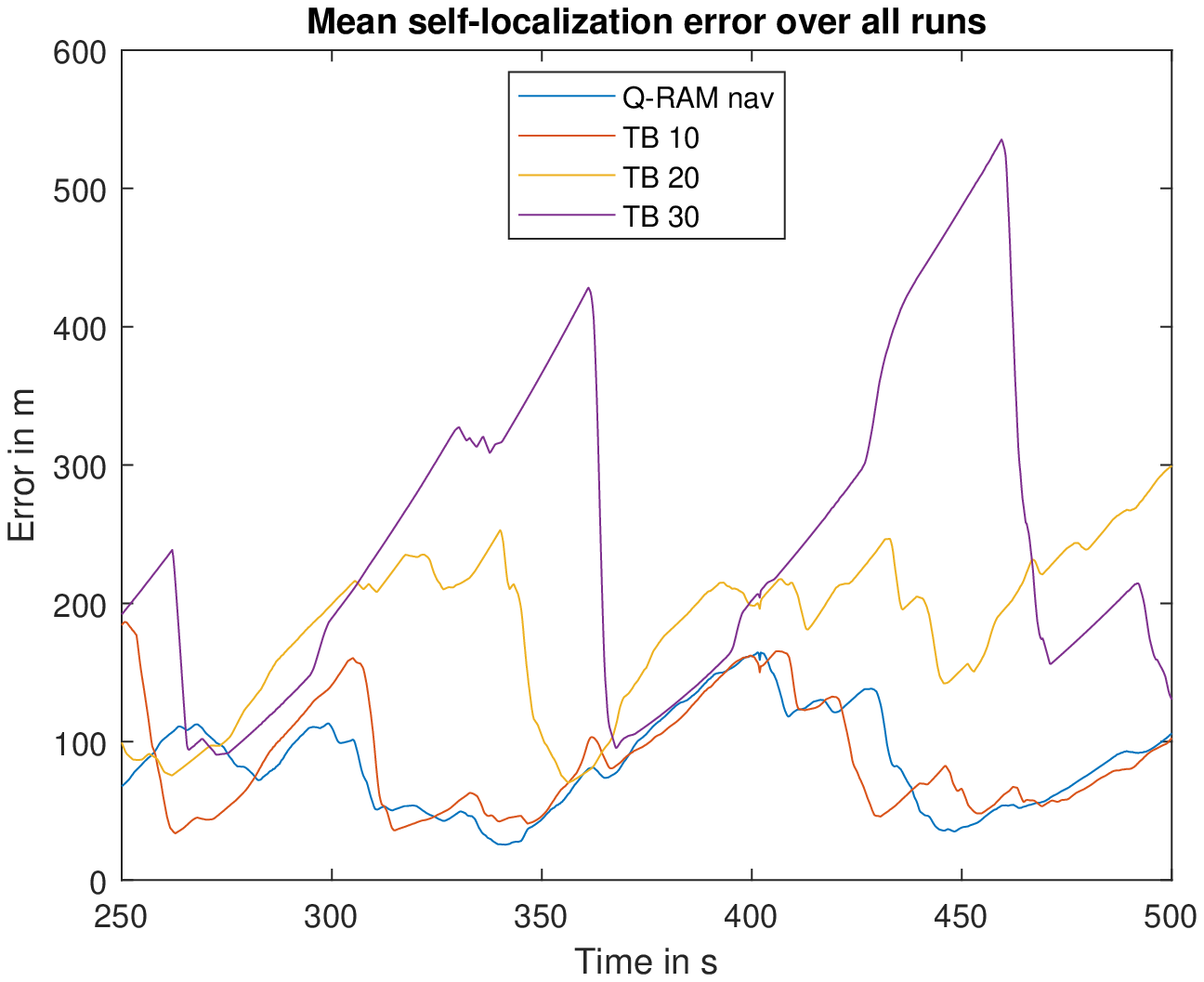}
	\caption{Mean self-localization errors for the different Q-RAM schemes (left) and the time balanced techniques (right).}\label{fig:selfTrackErr}
\end{figure}

Figure~\ref{fig:selfTrackErr} shows that all Q-RAM schemes should be preferred over the time balanced approaches in terms of self-localization quality, as they all have a similar performance that can be matched only by \emph{TB 10}. Notice however, that the tracking quality of \emph{TB 10} is by far the worst among the contestants.

\begin{figure}[htbp]
	\centering
	\includegraphics[width=0.45\textwidth]{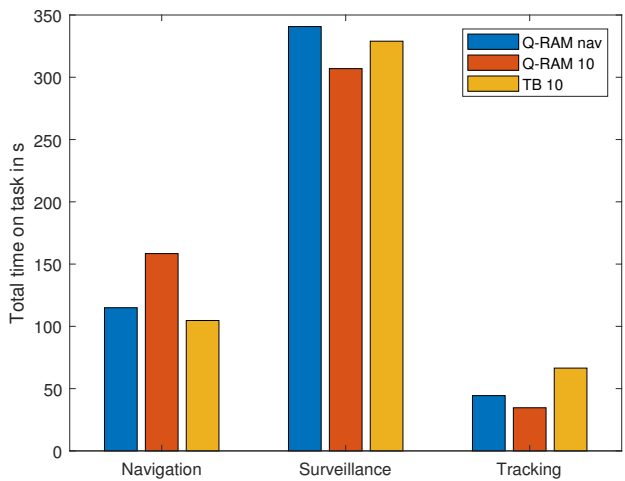}
	\caption{Resource usage for the different task types and allocation schemes.}\label{fig:resourceload}
\end{figure}

Additionally, Figure~\ref{fig:resourceload} highlights that compared with the algorithms with the best self-localization error, \emph{Q-RAM nav} uses the least resources for navigation and tracking, while achieving a considerably better tracking performance.

Although fixed high navigation update rates guarantee a good self-localization, they have a severe negative impact on the tracking performance even for comparably easy to manage scenarios like the one under investigation.
On the other hand, under fixed low navigation update rates the self-localization suffers, which also can significantly increase track errors in some situations.
Our solution excels in this trade-off by adapting the update rate to the current situation and mission requirements.

\section{Conclusion} \label{sec:conclusion}
The proposed localization method combined with the presented resource allocation scheme provides a reliable method for navigating even in hostile, GNSS-free environments while at the same time optimizing the overall system performance.

In particular, we have shown that it is likely possible to do self-localization on a radar by measuring the position of satellites in low earth orbit. Since this is a highly time demanding task, an adapted radar resource management is necessary to maintain overall system stability and performance.
Our method leads to a new concept of understanding the utility of tasks which, unlike search or track update tasks, do not have an inherent utility of their own but impact the quality of other tasks. The main advantage of this method is to use existing models of traditional tasks to calculate the utility of new tasks like localization -- but also calibration and synchronization -- with only a low development effort.
A potential downside is that for every configuration of such tasks a Q-RAM solution has to be calculated, which can lead to an increased computational load. However, as it is straight-forward to parallelize the algorithm, this problem can be easily overcome.

Future work is to consider a more realistic IMU model where error sources depend on movements of the aircraft and the gyroscope errors are included as well. This may lead to an additional resource load during or shortly after manoeuvres.
Additionally, the association problem has to be considered, with a more complete satellite catalogue. The Q-RAM should then be modified to reflect the miss-association probability such that it becomes a quality factor. Furthermore, refraction (leading to additional restrictions to the beam's minimum elevation), errors in the satellite database or propagation methods, surveillance induced track updates 
and more complex mission profiles (e.g.\ including classification, jamming and different targets)
could be considered.

\bibliography{mybibfile}

\end{document}